\documentclass[prb,aps,twocolumn,showpacs]{revtex4-1}

\usepackage{amsfonts,amsmath,bm}
\usepackage{graphicx}

\newcommand{\rl}{\rangle\!\langle}

\begin{document}

\title{Polaron contributions to the biexciton binding energies in
self-assembled quantum dots}

\author{Pawe{\l} Machnikowski}
 \email{Pawel.Machnikowski@pwr.wroc.pl} 
\affiliation{Institute of Physics, Wroc{\l}aw University of
Technology, 50-370 Wroc{\l}aw, Poland}

\begin{abstract}
The contribution to the biexciton binding
energy in quantum dots resulting from the interaction with
longitudinal optical phonons is estimated by performing the
configuration--interaction 
calculation of the few-particle states in a simple model of the
confining potential and including the phonon corrections by means of a
perturbation theory. It is found that the polaron contribution tends
to compensate the Coulomb-related biexciton shift (binding energy) and reduces
its value by several to even 30\%, depending on the material
parameters of the system.
\end{abstract}


\maketitle


The biexciton binding energy (also referred to as the biexciton shift) 
is one of the most important parameters
determining the optical properties of quantum dots (QDs). For instance, the
spectral shift between the exciton and biexciton transition makes it
possible to independently address various degrees of freedom of the
biexciton system which opens the way to a range of quantum optical
control schemes \cite{li03,stufler06}. The magnitude of this shift is
usually of the order of a few meV while its sign may be either positive
or negative \cite{hu90,rodt03b}.
Theoretically, the effects of Coulomb correlations in few-electron
QDs, underlying the biexciton shift, can be calculated using
a configuration-interaction scheme \cite{rontani06} either using a simple model of
confinement \cite{wojs96c,wojs96b,brasken00} or built upon more realistic wave
functions obtained within the multiple-band kp or pseudopotential
theory \cite{stier02}.

One factor that is missing from all the existing calculations is the
effect of the interaction between carriers and
longitudinal optical (LO) phonon which is of particular
importance in confined systems \cite{bawendi90,fomin98} and leads, among
other effects, to polaron shifts and resonant anticrossings
\cite{hameau02,obreschkow07,kaczmarkiewicz10}. According to theoretical 
calculations, the magnitude of these features may be as 
large as a few meV for a single confined 
exciton in a self-assembled QD \cite{verzelen02a,jacak03b}. Since this is
comparable to the biexciton binding energy it seems
interesting to study the phonon-related shift for a confined
biexciton. 

The purpose of this paper is to estimate the polaron contribution to
the biexciton binding energy in a QD. This will be done by calculating
the polaron correction to the exciton and biexciton states using the
perturbation theory. The few-particle states themselves will be
calculated within the usual configuration-interaction scheme with a
simple parabolic model of confinement. 

The quantum dot will be modeled by axially symmetric
harmonic oscillator potentials for electrons and holes with a strong
confinement in the growth 
direction. In order to mimic (in a simple way) the spatial charge separation
observed in many QD structures, the minima of the potentials for
electrons and holes are allowed to be displaced along the growth
direction and located at different points
$z_{\mathrm{e}}$ and $z_{\mathrm{h}}$, respectively. 
The single particle Hamiltonian is then
\begin{eqnarray*}
\lefteqn{H^{(\mathrm{e/h})}=}\\
&&-\frac{\hbar^{2}}{2m_{\mathrm{e/h}}}\Delta
+\frac{1}{2}m_{\mathrm{e/h}}\omega_{\mathrm{e/h}}^{2}r_{\bot}^{2}
+\frac{1}{2}m_{\mathrm{e/h}}\tilde{\omega}_{\mathrm{e/h}}^{2}(z-z_{\mathrm{e/h}})^{2},
\end{eqnarray*}
where ``e/h'' corresponds to the electron and hole wave functions,
$m_{\mathrm{e/h}}$ are the effective masses, and
$\omega_{\mathrm{e/h}}$, $\tilde{\omega}_{\mathrm{e/h}}$
are the harmonic oscillator frequencies for the confinement potential in
the $xy$ plane and along the $z$ axis, respectively, with 
$\tilde{\omega}_{\mathrm{e/h}}\gg\omega_{\mathrm{e/h}}$.
The motion along the $z$ axis (growth direction) will be
restricted to the ground state. 
It will be assumed that the confinement width of a single
(non-interacting) particle is the same for the electrons and holes,
\begin{displaymath}
l_{0}=\sqrt{\frac{\hbar}{m_{\mathrm{e}}\omega_{\mathrm{e}}}}
=\sqrt{\frac{\hbar}{m_{\mathrm{h}}\omega_{\mathrm{h}}}},\quad
l_{z}=\sqrt{\frac{\hbar}{m_{\mathrm{e}}\tilde{\omega}_{\mathrm{e}}}}
=\sqrt{\frac{\hbar}{m_{\mathrm{h}}\tilde{\omega}_{\mathrm{h}}}},
\end{displaymath}
for the in-plane and axial confinement, respectively.

The calculations will be performed in the basis of single-particle
wave functions
\begin{equation}\label{wavefun}
\Psi_{nm}^{\mathrm{(e/h)}}(\bm{r})
=\Phi(z)\psi_{n,m}^{\mathrm{(e/h)}}(\bm{r}_{\bot}),
\end{equation}
where 
\begin{displaymath}
\Phi(z)=\frac{1}{\pi^{1/4}l_{z}^{1/2}}\exp\left[-\frac{(z-z_{\mathrm{e/h}})^{2}}{2l_{z}^{2}}\right]
\end{displaymath}
is the wave function in the $z$ direction and
\begin{eqnarray*}
\lefteqn{\psi_{n,m}^{(\mathrm{e/h})}(\bm{r}_{\bot})=}\\
&&\frac{1}{l_{\mathrm{e/h}}}\sqrt{\frac{2n!}{(|m|+n)!}}
e^{-r_{\bot}^{2}/(2l_{\mathrm{e/h}}^{2})}
\left(\frac{r_{\bot}}{l_{\mathrm{e/h}}}  \right)^{|m|}
\mathcal{L}_{n}^{|m|}\left( \frac{r_{\bot}^{2}}{l_{\mathrm{e/h}}^{2}} \right)
\end{eqnarray*}
is the Fock--Darwin wave function of the
2-dimensional harmonic oscillator \cite{jacak98a},
where $\mathcal{L}_{n}^{|m|}(x)$ is the Laguerre polynomial.
Here 
$n=0,1,\ldots$ is the radial quantum number and $m=0,\pm 1,\ldots$ 
is the angular momentum quantum number.
The length parameters $l_{\mathrm{e}}$ and $l_{\mathrm{h}}$ are
determined by minimizing the variational exciton ground state
\begin{eqnarray*}
\lefteqn{E^{(\mathrm{var})} =} \\
&&\int d^{3}r \int d^{3}r'
\Psi_{00}^{\mathrm{(e)}*}(\bm{r}) \Psi_{00}^{\mathrm{(h)}*}(\bm{r'}) \\
&&\times\left[
H^{(\mathrm{e})}(\bm{r})+H^{(\mathrm{h})}(\bm{r}')- V(\bm{r},\bm{r}')  \right]
\Psi_{00}^{\mathrm{(h)}}(\bm{r}') \Psi_{00}^{\mathrm{(e)}}(\bm{r}),
\end{eqnarray*}
where 
\begin{displaymath}
V(\bm{r},\bm{r}')= 
\frac{e^{2}}{4\pi\varepsilon_{0}\varepsilon_{\mathrm{s}}}
\frac{1}{|\bm{r}-\bm{r}'|},
\end{displaymath}
$\varepsilon_{0}$ is the vacuum permittivity, and
$\varepsilon_{\mathrm{s}}$ is the static relative dielectric constant.
In the following, we will use a single
Greek index $\alpha,\beta,\ldots$ to represent the pair of quantum
numbers $(n,m)$. 

The confined few-particle electron--hole subsystem is then described by the
Hamiltonian 
\begin{eqnarray}
H_{\mathrm{e-h}} & = & 
\sum_{\alpha\beta\sigma}\epsilon^{(\mathrm{e})}_{\alpha\beta}
a_{\alpha\sigma}^{\dag}a_{\beta\sigma}
+\sum_{\alpha\beta\sigma}\epsilon^{(\mathrm{h})}_{\alpha\beta}
h_{\alpha\sigma}^{\dag}h_{\beta\sigma} \nonumber \\
&&+\sum_{\alpha\beta\alpha'\beta'}\sum_{\sigma\sigma'}
V_{\mathrm{ee}}(\alpha,\beta;\alpha',\beta')
a_{\alpha\sigma}^{\dag} a_{\beta\sigma'}^{\dag}
a_{\beta'\sigma'}a_{\alpha'\sigma} \nonumber \\
&& +\sum_{\alpha\beta\alpha'\beta'}\sum_{\sigma\sigma'}
V_{\mathrm{hh}}(\alpha,\beta;\alpha',\beta')
h_{\alpha\sigma}^{\dag} h_{\beta\sigma'}^{\dag}
h_{\beta'\sigma'}h_{\alpha'\sigma} \nonumber \\
&& -\sum_{\alpha\beta\alpha'\beta'}\sum_{\sigma\sigma'}
V_{\mathrm{eh}}(\alpha,\beta;\alpha',\beta')
a_{\alpha\sigma}^{\dag} h_{\beta\sigma'}^{\dag}
h_{\beta'\sigma'}a_{\alpha'\sigma},
\label{H-e-h}
\end{eqnarray}
where $a_{\alpha\sigma},a_{\alpha\sigma}^{\dag}$ and
$h_{\alpha\sigma},h_{\alpha\sigma}^{\dag}$ are the annihilation and
creation operators for electrons and holes, respectively, corresponding to the
states described by the wave functions given by Eq.~\eqref{wavefun}
and spin orientation $\sigma$,
\begin{displaymath}
\epsilon^{(\mathrm{e/h})}_{\alpha\beta}
=\int d^{3}r \Psi_{\alpha}^{(\mathrm{e/h})*}(\bm{r}) 
H^{(\mathrm{e/h})}(\bm{r}) \Psi_{\beta}^{(\mathrm{e/h})}(\bm{r}) 
\end{displaymath}
and
\begin{eqnarray*}
\lefteqn{V_{ij}(\alpha,\beta;\alpha',\beta')=} \\
&&\int d^{3}r \int d^{3}r'
\Psi_{\alpha}^{(i)*}(\bm{r}) \Psi_{\beta}^{(j)*}(\bm{r'})
 V(\bm{r},\bm{r}')
\Psi_{\beta'}^{(j)}(\bm{r}') \Psi_{\alpha'}^{(i)}(\bm{r}),
\end{eqnarray*}
with $i,j=\mathrm{e,h}$.

The interaction between the carriers and LO
phonons is described by the Fr\"ohlich Hamiltonian \cite{jacak03b} 
\begin{eqnarray}
H_{\mathrm{c-ph}}& = & 
e\sqrt{\frac{\hbar\Omega}{2V\epsilon_{0}\tilde{\epsilon}}}
\sum_{\bm{k}}\frac{1}{k}
\sum_{\alpha\beta\sigma} \left[ 
\mathcal{F}^{(\mathrm{h})}_{\alpha\beta}h^{\dag}_{\alpha\sigma}h_{\beta\sigma}(\bm{k})\right. 
\nonumber\\
&&\left.
-\mathcal{F}^{(\mathrm{e})}_{\alpha\beta}a^{\dag}_{\alpha\sigma}a_{\beta\sigma}(\bm{k})
\right]  \left( b_{\bm{k}}+b_{-\bm{k}}^{\dag} \right),
\label{c-ph}
\end{eqnarray}
where
$\tilde{\varepsilon}=(1/\varepsilon_{\mathrm{s}}-1/\varepsilon_{\infty})^{-1}$
is the effective dielectric constant ($\varepsilon_{\infty}$ is the
high-frequency relative dielectric constant), $\Omega$ is the frequency
of the LO phonons (assumed constant in the narrow range of the
relevant wave vectors), $b_{\bm{k}}$ and $b_{\bm{k}}^{\dag}$ are the
annihilation and creation operators for the LO phonon with the wave
vector $\bm{k}$, 
$V$ is the normalization volume for phonons,
and the form factors are given by
\begin{displaymath}
\mathcal{F}^{(\mathrm{e/h})}_{\alpha\beta}(\bm{k})=
\int d^{3}r \Psi_{\alpha}^{(\mathrm{e/h})*}(\bm{r}) 
e^{i\bm{k}\cdot\bm{r}}\Psi_{\beta}^{(\mathrm{e/h})}(\bm{r}).
\end{displaymath}

First, the Coulomb part of the biexciton shift $\Delta E_{\mathrm{C}}$
is determined. To this end, the Hamiltonian \eqref{H-e-h} is
diagonalized in the truncated basis of one-pair states
yielding the exciton energies $E_{n}^{(1)}$ and the
corresponding eigenstates
\begin{equation}
|\mathrm{X}n\rangle = \sum_{\alpha\beta}c^{(n)}_{\alpha\beta}
a_{\alpha\sigma}^{\dag}h_{\beta\sigma'}^{\dag}|0\rangle,
\label{1-p}
\end{equation}
where $|0\rangle$ is the empty dot state. The subtle
fine structure effects are not included in the model, so the states
are degenerate with respect to spin configurations. As both the
Coulomb interaction and the carrier-phonon couping conserves separately
the electron and hole spins and we are interested in the ground state
only it is sufficient to take into account the two-pair 
states with singlet spin configurations. The energies of
the biexciton states are therefore found by diagonalizing the
Hamiltonian $H_{\mathrm{e-h}}$ in the basis of two-pair spin-singlet
states, from which 
one finds the energies $E_{n}^{(2)}$ and the corresponding
eigenstates
\begin{eqnarray}
|\mathrm{XX}n\rangle & = &  \sum_{\alpha\ge\alpha'}\sum_{\beta\beta'}
\eta_{\alpha\alpha'}\eta_{\beta\beta'}
C^{(n)}_{\alpha\alpha'\beta\beta'} \nonumber \\
&&\times\frac{a_{\alpha\uparrow}^{\dag}a_{\alpha'\downarrow}^{\dag}
+a_{\alpha'\uparrow}^{\dag}a_{\alpha\downarrow}^{\dag}}{2}
\frac{h_{\beta\uparrow}^{\dag}h_{\beta'\downarrow}^{\dag}
+h_{\beta'\uparrow}^{\dag}h_{\beta\downarrow}^{\dag}}{2}
|0\rangle,
\label{2-p}
\end{eqnarray}
where $\uparrow$ and $\downarrow$ denote spin orientations,
$\eta_{\alpha\beta}=(1-\sqrt{2})\delta_{\alpha\beta}+\sqrt{2}$, and
comparing the labels $\alpha,\beta$ is performed in the sense of an
arbitrary but fixed ordering of the single-particle states.
The Coulomb part of the ground state biexciton shift is then simply
\begin{displaymath}
\Delta E_{\mathrm{C}}=E^{(2)}_{0}-2E^{(1)}_{0}.
\end{displaymath}

Next, one has to compute the phonon-induced corrections to the exciton
and biexciton states. Even though the polaron shifts for single
carriers can be quite large, those appearing for globally neutral
exciton states are smaller due to partial cancellation of the
electron--phonon and hole--phonon interactions. Therefore, a
reliable estimate of the polaron shift in the cases considered here
can be obtained by means of the second order perturbation theory. The
calculation is done again separately for the exciton and biexciton cases.

In the case of an exciton, one rewrites the
carrier-phonon interaction in terms of the one-pair eigenstates
[Eq.~\eqref{1-p}] (for an arbitrary, fixed spin configuration), 
\begin{displaymath}
H_{\mathrm{c-ph}}^{(1)}=
e\sqrt{\frac{\hbar\Omega}{2V\epsilon_{0}\tilde{\epsilon}}}
\sum_{\bm{k}\sigma}\frac{1}{k}
\sum_{nm}\mathcal{G}^{(1)}_{nm}(\bm{k})
|\mathrm{X}n\rl \mathrm{X}m|,
\end{displaymath}
where 
\begin{equation}
\mathcal{G}^{(1)}_{nm}(\bm{k})=
\sum_{\alpha\alpha'\beta\beta'}
c^{(n)*}_{\alpha\beta}c^{(m)}_{\alpha'\beta'}
\left[ \delta_{\alpha\alpha'}\mathcal{F}^{(\mathrm{h})}_{\beta\beta'}
-\delta_{\beta\beta'}\mathcal{F}^{(\mathrm{e})}_{\alpha\alpha'}
\right].
\label{G1}
\end{equation}
In an analogous manner, for the biexciton, one writes
\begin{displaymath}
H_{\mathrm{c-ph}}^{(2)}=
e\sqrt{\frac{\hbar\Omega}{2V\epsilon_{0}\tilde{\epsilon}}}
\sum_{\bm{k}\sigma}\frac{1}{k}
\sum_{nm}\mathcal{G}^{(2)}_{nm}(\bm{k})
|\mathrm{XX}n\rl \mathrm{XX}m|,
\end{displaymath}
where
\begin{eqnarray}
\mathcal{G}^{(2)}_{nm}(\bm{k})& = & 
\frac{1}{2}
\sum_{\mu'\ge\nu'}\sum_{\mu\ge\nu}\sum_{\kappa\lambda}
C^{(n)*}_{\kappa\lambda\mu'\nu'}C^{(m)}_{\kappa\lambda\mu\nu}
 \eta_{\mu\nu}\eta_{\mu'\nu'}\nonumber\\
&&\times\left[
\mathcal{F}^{\mathrm{(h)}}_{\mu'\mu}(\bm{k})\delta_{\nu'\nu}
+\mathcal{F}^{\mathrm{(h)}}_{\nu'\mu}(\bm{k})\delta_{\mu'\nu}\right.
\nonumber\\
&&\left.
+\mathcal{F}^{\mathrm{(h)}}_{\nu'\nu}(\bm{k})\delta_{\mu'\mu}
+\mathcal{F}^{\mathrm{(h)}}_{\mu'\nu}(\bm{k})\delta_{\nu'\mu}
 \right]\nonumber\\
&&-\frac{1}{2}
\sum_{\mu'\ge\nu'}\sum_{\mu\ge\nu}\sum_{\kappa\lambda}
C^{(n)*}_{\mu'\nu'\kappa\lambda}C^{(m)}_{\mu\nu\kappa\lambda}
 \eta_{\mu\nu}\eta_{\mu'\nu'}\nonumber\\
&&\times\left[
\mathcal{F}^{\mathrm{(e)}}_{\mu'\mu}(\bm{k})\delta_{\nu'\nu}
+\mathcal{F}^{\mathrm{(e)}}_{\nu'\mu}(\bm{k})\delta_{\mu'\nu}\right. \nonumber\\
&&\left.
+\mathcal{F}^{\mathrm{(e)}}_{\nu'\nu}(\bm{k})\delta_{\mu'\mu}
+\mathcal{F}^{\mathrm{(e)}}_{\mu'\nu}(\bm{k})\delta_{\nu'\mu}
 \right].
\label{G2}
\end{eqnarray}
The phonon-induced correction (polaron shift) to the ground state
energy is then given by
\begin{equation}
\delta E^{(i)}=
-\frac{e^{2}\hbar\Omega}{16\pi^{3}\varepsilon_{0}\tilde{\varepsilon}}
\sum_{n}\int \frac{d^{3}k}{k^{2}}
\frac{|\mathcal{G}^{(i)}(\bm{k})|^{2}}{
E_{n}^{(i)}-E_{0}^{(i)}+\hbar\Omega},
\label{shift}
\end{equation}
where $i=1,2$ refers to the exciton and biexciton cases, respectively.
The phonon correction to the biexciton shift is
\begin{equation}
\Delta E_{\mathrm{ph}}=\delta E^{(2)}-2\delta E^{(1)}.
\label{biex-ph}
\end{equation}
The total biexciton shift is 
$\Delta E=\Delta E_{\mathrm{C}}+\Delta E_{\mathrm{ph}}$. 

\begin{table}[tb]
  \centering
  \begin{tabular}{lcc}
\hline
    & GaAs & CdTe \\
\hline
   $\hbar\Omega$ & 36~meV& 21~meV \\
   $\epsilon_{\mathrm{s}}$ & 12.9 & 10.2 \\
   $\epsilon_{\infty}$ & 10.86 & 7.2 \\
   $m_{\mathrm{e}}$ & $0.066m_{0}$ & $0,09m_{0}$ \\
   $m_{\mathrm{h}}$ & $0.3m_{0}$ & $0.8m_{0}$\\
\hline
  \end{tabular}
  \caption{Material parameters used in the calculations. $m_{0}$ is
    the free electron mass.}
  \label{tab:param}
\end{table}

In the numerical calculations, two sets
of parameters will be used, representing two groups of material
systems: GaAs (a moderately polar III-V compound) and CdTe 
(a II-VI compound with a more polar character) (see
Tab.~\ref{tab:param}). The numerical computations are 
performed using the basis of 6 electron and hole shells ($2|m|+n\le
5$), that is, 21 electron and hole wave functions. This is sufficient
to assure the convergence of numerical results within $\sim 0.1$~meV
of possible error, which is sufficient in view of the much larger
magnitude of the discussed effect (a few meV).

\begin{figure}[tb]
\begin{center}
\includegraphics[width=85mm]{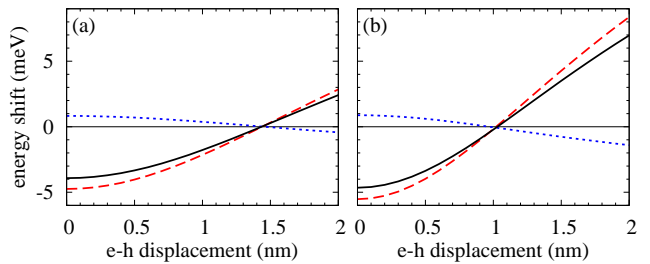}
\end{center}
\caption{\label{fig:GaAs}The  biexciton shift (black
  solid lines) for a   GaAs--like QD system, as well as its Coulomb (red dashed lines) and phonon
  (blue dotted lines) components for the two lateral QD sizes
  characterized by $\hbar\omega_{\mathrm{e}}=40$~meV (a) and 70~meV
  (b). In both cases $l_{z}=0.2l_{0}$.}
\end{figure}

The results for
the first set of parameters are shown in Fig.~\ref{fig:GaAs}, where the dependence
of the biexciton shift on the e-h displacement
$D=z_{\mathrm{e}}-z_{\mathrm{h}}$ is shown for two lateral QD
sizes. Although the polaron contribution does not essentially change
the qualitative behavior of the biexciton shift it constitutes an important
correction to the overall value. For this material system, the phonon-induced contribution
amounts to roughly 15-17\% of the Coulomb part over the whole range of
e-h displacements $D$ studied here, except for the region where the binding
energy changes sign and this ratio is poorly defined. The
relative value of the polaron part is very similar for both QD sizes
even though the absolute magnitude of the total binding energy is
larger for the smaller dot (larger $\hbar\omega_{\mathrm{e}}$). It turns out
that the phonon correction has the opposite sign to the Coulomb
contribution and, therefore, reduces the absolute value of the binding
energy. Interestingly, both these contributions change sign at almost
the same value of $D$. 

\begin{figure}[tb]
\begin{center}
\includegraphics[width=85mm]{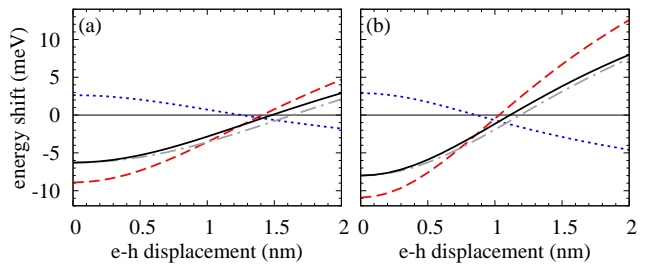}
\end{center}
\caption{\label{fig:CdTe}As in Fig. \ref{fig:GaAs} and for the same
  two values of $\hbar\omega_{\mathrm{e}}$ but for a CdTe-like
  system. The gray dash-dotted line shows the biexciton shift with the
correction due to the axial shift.}
\end{figure}

In Fig.~\ref{fig:CdTe}, analogous results for a more strongly polar
system are shown. The magnitude of both the contributions to the
binding energy is now much larger. Also the role of the phonon
correction becomes more 
important and its value reaches almost 30\% for both dot
sizes. Again, the signs of the two contributions are opposite over
most of the parameter range studied. In this case, the sign change
still appears in the same region for both contributions but not as
closely as in the previous case.

In the presented discussion, the wave functions along the growth
direction were restricted to the ground state, that is, the charge
distribution was supposed to be rigid with respect to axial
shifts. However, for $D>0$, the Coulomb attraction will
lead to an axial shift of the electron and hole charge distributions,
which will contribute to the biexciton shift. Accounting for this effect
requires extending the computational basis by including the first
excited state wave
functions along $z$. By
virtue of the Ritz theorem, the same goal (to the leading order) can
be achieved by a variational minimization of the system energy with
respect to the positions of the centers of the wave functions along
$z$ for a given distance $D$ between the minima of the electron and
hole confinement potentials. The gray dash-dotted line in
Fig.~\ref{fig:CdTe} shows the biexciton shift for the CdTe-like system
including the additional contribution from the axial shift obtained by
such a procedure (with fixed $l_{\mathrm{e/h}}$ found earlier). The
correction reaches almost 1~meV at $D=2$~nm, which is noticeable,
although not particularly large compared to the values without the
correction. For a GaAs-like system, the correction does not exceed
0.2~meV and is therefore relatively much smaller. The
shift of the charge distributions does not exceed 0.04~nm and 0.1~nm for
the GaAs and CdTe system, respectively. Therefore,
the correction to the phonon part, which is only sensitive to the
geometry of the wave functions, will reach 0.03~meV and
0.25~meV, respectively, as can be deduced from the slopes of the
phonon curves in Figs.~\ref{fig:GaAs} and~\ref{fig:CdTe}. In both
cases, this corrections is rather 
small compared to the overall values of the biexciton shift.

Finally, it can be argued that the results presented above are
in a way nontrivial, at least in the sense that they cannot be
reproduced using a simple model. Indeed, if one considers only exciton
and biexciton wave functions in a product (Hartree) form and takes into
account that the wave functions for the two electrons (or for the two
holes) that differ only by their spin orientation must be the same
then the resulting biexciton shift must be positive even if the
single-particle wave functions in the biexciton state are allowed to
be different from those in the exciton state. Therefore, the
negative value of the shift (positive binding energy) is a purely
correlation-induced effect. 

For the phonon effect, it seems impossible to draw such a general
conclusion since the polaron correction depends strongly on the charge
cancellation and hence on the exact shape of the wave functions which
can differ between the exciton and biexciton cases. However, if one
restricts the discussion to a single-orbital model in which only one
wave function is available for electrons and one for holes (the same
in the exciton and biexciton configurations) then it can be seen from
Eqs.~\eqref{G1} and~\eqref{G2} that
\begin{displaymath}
\mathcal{G}^{(2)}(\bm{k})=2\left[ \mathcal{F}^{(\mathrm{h})}(\bm{k})
-\mathcal{F}^{(\mathrm{e})}(\bm{k})   \right]=2\mathcal{G}^{(1)}(\bm{k}),
\end{displaymath}
where the indices of the form factors have been omitted. As there is
only one state, only one term ($n=0$) remains in Eq.~\eqref{shift}
and, clearly, $\delta E^{(2)}=4\delta E^{(1)}$. Therefore, according
to Eq.~\eqref{biex-ph},
$\Delta E_{\mathrm{ph}}=2\delta E^{(1)}<0$, since the ground state
polaron shift is negative. Clearly, this result is opposite to that
obtained from the full model, where the phonon correction to the
biexciton shift is positive for small $D$. Therefore, exact shape of
the wave functions, including admixtures of higher single particle
states is essential for correct modeling. 

In summary, the LO phonon-related (polaronic)
contribution to the biexciton shift (biexciton binding energy) in
quantum dots has been calculated. 
It turns out that this correction is about 15\% 
of the dominating Coulomb contribution in a typical III-V material 
and reaches almost 30\% of the Coulomb term for a more strongly polar
II-VI system. The phonon correction usually has an opposite sign to the
Coulomb part and, hence, reduces the binding energy of a confined
biexciton. 

This work was supported by the TEAM programme of the Foundation for
Polish Science co-financed from the European Regional Development
Fund.


\begin{thebibliography}{10}

\bibitem{li03}
X. Li, Y. Wu, D. Steel, D. Gammon, T. Stievater, D. Katzer, D. Park, C.
  Piermarocchi, and L. Sham, Science {\bf 301},  809  (2003).

\bibitem{stufler06}
S. Stufler, P. Machnikowski, P. Ester, M. Bichler, V.~M. Axt, T. Kuhn, and A.
  Zrenner, Phys. Rev. B {\bf 73},  125304  (2006).

\bibitem{hu90}
Y.~Z. Hu, S.~W. Koch, M. Lindberg, N. Peyghambarian, E.~L. Pollock, and F.~F.
  Abraham, Phys. Rev. Lett. {\bf 64},  1805  (1990).

\bibitem{rodt03b}
S. Rodt, R. Heitz, A. Schliwa, R.~L. Sellin, F. Guffarth, and D. Bimberg, Phys.
  Rev. B {\bf 68},  035331  (2003).

\bibitem{rontani06}
M. Rontani, C. Cavazzoni, D. Bellucci, and G. Goldoni, J. Chem. Phys. {\bf
  124},  124102  (2006).

\bibitem{wojs96c}
A. W{\'o}js and P. Hawrylak, Acta Phys. Polon. A {\bf 90},  1108  (1996).

\bibitem{wojs96b}
A. W{\'o}js and P. Hawrylak, Solid State Commun. {\bf 100},  487  (1996).

\bibitem{brasken00}
M. Brasken, M. Lindberg, D. Sundholm, and J. Olsen, Phys. Rev. B {\bf 61},
  7652  (2000).

\bibitem{stier02}
O. Stier, R. Heitz, A. Schliwa, and D. Bimberg, Phys. Stat. Sol. (a) {\bf 190},
   477  (2002).

\bibitem{bawendi90}
M.~G. Bawendi, W.~L. Wilson, L. Rothberg, P.~J. Carroll, T.~M. Jedju, M.~L.
  Steigerwald, and L.~E. Brus, Phys. Rev. Lett. {\bf 65},  1623  (1990).

\bibitem{fomin98}
V.~M. Fomin, V.~N. Gladilin, J.~T. Devreese, E.~P. Pokatilov, S.~N. Balaban,
  and S.~N. Klimin, Phys. Rev. B {\bf 57},  2415  (1998).

\bibitem{hameau02}
S. Hameau, J.~N. Isaia, Y. Guldner, E. Deleporte, O. Verzelen, R. Ferreira, G.
  Bastard, J. Zeman, and J.~M. G\'erard, Phys. Rev. B {\bf 65},  085316
  (2002).

\bibitem{obreschkow07}
D. Obreschkow, F. Michelini, S. Dalessi, E. Kapon, and M.-A. Dupertuis, Phys.
  Rev. B {\bf 76},  035329  (2007).

\bibitem{kaczmarkiewicz10}
P. Kaczmarkiewicz and P. Machnikowski, Phys. Rev. B {\bf 81},  115317  (2010).

\bibitem{verzelen02a}
O. Verzelen, R. Ferreira, and G. Bastard, Phys. Rev. Lett. {\bf 88},  146803
  (2002).

\bibitem{jacak03b}
L. Jacak, P. Machnikowski, J. Krasnyj, and P. Zoller, Eur. Phys. J. D {\bf 22},
   319  (2003).

\bibitem{jacak98a}
L. Jacak, P. Hawrylak, and A. Wojs, {\em Quantum Dots} (Springer Verlag,
  Berlin, 1998).

\end{thebibliography}

\end{document}